\documentstyle[12pt]{article}
\def\lsim{\:\raisebox{-0.5ex}{$\stackrel{\textstyle<}{\sim}$}\:}
\def\gsim{\:\raisebox{-0.5ex}{$\stackrel{\textstyle>}{\sim}$}\:}
\def\opc{\hbox{{\sl op. cit.} }}
\newcommand {\ignore}[1]{}

\newcommand{\noi}{\noindent}
\newcommand{\bc}{\begin{center}}
\newcommand{\ec}{\end{center}}

\def\ifmath#1{\relax\ifmmode #1\else $#1$\fi}
\def\3quarter{{\textstyle{3 \over 4}}}

\def\ra{\rightarrow}

\def\lf{\leaders\hbox to 1em{\hss.\hss}\hfill}

\def\21{$SU(2) \ot U(1)$}
\def\321{$SU(3) \ot SU(2) \ot U(1)$}
\def\ne{\hbox{$\nu_e$ }}
\def\nm{\hbox{$\nu_\mu$ }}
\def\nt{\hbox{$\nu_\tau$ }}
\def\ns{\hbox{$\nu_{sterile}$ }}

\def\ns{\hbox{$\nu_S$ }}

\def\O{\hbox{$\cal O$ }}

\def\mnt{\hbox{$m_{\nu_\tau}$ }}

\def\etal{\hbox{\it et al., }}
\def\neu{\hbox{neutrino }}

\def\neus{\hbox{neutrinos }}

\def\eq#1{{eq. (\ref{#1})}}

\def\abs#1{\left| #1\right|}

\def\lsim{\raise0.3ex\hbox{$\;<$\kern-0.75em\raise-1.1ex\hbox{$\sim\;$}}}
\def\gsim{\raise0.3ex\hbox{$\;>$\kern-0.75em\raise-1.1ex\hbox{$\sim\;$}}}
\def\bel{\begin{letter}}
\def\eel{\end{letter}}
\def\beq{\begin{equation}}
\def\eeq{\end{equation}}
\def\bef{\begin{figure}}
\def\eef{\end{figure}}
\def\bet{\begin{table}}
\def\eet{\end{table}}
\def\bea{\begin{eqnarray}}
\def\ba{\begin{array}}
\def\ea{\end{array}}
\def\bi{\begin{itemize}}
\def\ei{\end{itemize}}
\def\ben{\begin{enumerate}}
\def\een{\end{enumerate}}
\def\ra{\rightarrow}
\def\ot{\otimes}
\def\eea{\end{eqnarray}}
\def\apj#1#2#3{          {\it Astrophys. J. }{\bf #1} (19#2) #3}

\def\ib#1#2#3{           {\it ibid. }{\bf #1} (19#2) #3}
\def\nat#1#2#3{          {\it Nature }{\bf #1} (19#2) #3}
\def\nps#1#2#3{          {\it Nucl. Phys. B (Proc. Suppl.) }
                         {\bf #1} (19#2) #3}
\def\np#1#2#3{           {\it Nucl. Phys. }{\bf #1} (19#2) #3}
\def\pl#1#2#3{           {\it Phys. Lett. }{\bf #1} (19#2) #3}
\def\pr#1#2#3{           {\it Phys. Rev. }{\bf #1} (19#2) #3}

\def\prl#1#2#3{          {\it Phys. Rev. Lett. }{\bf #1} (19#2) #3}

\def\zp#1#2#3{           {\it Zeit. fur Physik }{\bf #1} (19#2) #3}

\def\n.c.#1#2#3{         {\it Nuovo Cim. }{\bf #1} (19#2) #3}
\def\r.n.c.#1#2#3{       {\it Riv. del Nuovo Cim. }{\bf #1} (19#2) #3}
\def\sjnp#1#2#3{         {\it Sov. J. Nucl. Phys. }{\bf #1} (19#2) #3}

\def\mpl#1#2#3{          {\it Mod. Phys. Lett. }{\bf #1} (19#2) #3}

\def\ppnp#1#2#3{           {\it Prog. Part. Nucl. Phys. }{\bf #1} (19#2) #3}

\def\ijmp#1#2#3{           {\it Int. J. Mod. Phys. }{\bf #1} (19#2) #3}

\relax
\begin{document}
\begin{titlepage}
\begin{center}
\hfill{FTUV/94-46}\\
\hfill{IFIC/94-41}\\
\vskip 0.3cm
{\Large \bf RECONCILING COLD DARK MATTER
WITH COBE/IRAS PLUS SOLAR AND ATMOSPHERIC
NEUTRINO DATA }\\
\vskip 1.0cm
{\large \bf ANJAN S. JOSHIPURA}
\footnote{Permanent address: Theory Group,
 Physical Research Lab., Ahmedabad, India. E-mail 16444::JOSHIPUR}
and
{\large \bf JOS\'E W. F. VALLE}
\footnote{E-mail VALLE at vm.ci.uv.es or 16444::VALLE}\\
\vskip .5cm
{\it Instituto de F\'{\i}sica Corpuscular - C.S.I.C.\\
Departament de F\'{\i}sica Te\`orica, Universitat de Val\`encia\\
46100 Burjassot, Val\`encia, SPAIN}\\
\end{center}

\begin{abstract}
\baselineskip=12pt
{
We present a model where an unstable MeV Majorana tau
\neu can naturally reconcile the cold dark matter model
(CDM) with cosmological observations of large and small
scale density fluctuations and, simultaneously, with data
on solar and atmospheric neutrinos. The solar \neu deficit
is explained through long wavelength, so-called
{\sl just-so} oscillations involving conversions of \ne
into both \nm and a sterile species \ns, while atmospheric
\neu data are explained through \nm to \ne conversions.
Future long baseline \neu oscillation experiments, as
well as some reactor experiments will test this hypothesis.
The model is based on the spontaneous violation
of a global lepton number symmetry at the weak scale.
This symmetry plays a key role in generating the
cosmologically required decay of the \nt with lifetime
$\tau_{\nu_\tau} \sim 10^2 - 10^4$ seconds, as well
as the masses and oscillations of the three light
\neus \ne, \nm and \ns required in order to account for solar
and atmospheric \neu data. It also leads to the
invisibly decaying higgs signature that can be
searched at LEP and future particle colliders.
}
\end{abstract}

\end{titlepage}

\section{Introduction}

A tau neutrino with a mass in the MeV range is an
interesting possibility to consider for two different
reasons. On experimental side such a neutrino is within
the range of the detectability, for example at a tau-charm
factory \cite{jj,tcf1}. On the other hand, if such neutrino
decays  before the matter dominance epoch, its decay
products could then add energy to the radiation thereby
delaying the time at which the matter and radiation
contributions to the energy density of the universe
become equal. Such delay would allow one to reduce
the density fluctuations at the smaller scales \cite{latedecay}
purely within the standard cold dark matter scenario \cite{cdm},
and could reconcile the large scale fluctuations observed by
COBE \cite{cobe} with the earlier observations such as those
of IRAS \cite{iras} on the fluctuations at smaller scales.
An MeV  $\nu_{\tau}$ may, however, conflict with the
big-bang nucleosynthesis picture \cite{nucl1}.
This conflict can be avoided in two ways.

If the tau neutrino has a strong coupling to a light
spin zero Goldstone boson - called majoron and denoted $J$ -
with a typical strength $g_J \gsim 10^{-4}$ , then the
annihilation of $\nu_{\tau}$ pairs to majorons could reduce
their number density sufficiently so as to be consistent with
the nucleosynthesis bound \cite{ma1}. In spite of this reduction,
subsequent $\nu_{\tau}$ decays could increase  the energy
density of the radiation enough to reconcile COBE with IRAS data.
This possibility has been recently investigated \cite{ma1}
in the context of a specific doublet majoron
model \cite{ma2}, where an upper bound on the required \nt
life of $10^{6}$ seconds was set.

Alternatively, even if the majoron coupling to the tau
neutrino is not so strong it may be possible to reconcile
the nucleosynthesis constraints with the MeV $\nu_{\tau}$
hypothesis if its decay involves the electron neutrino.
The decay \ne could be captured by neutrons so as to reduce the
resulting yield of primordial helium. In this case,
a $\nu_{\tau}$ with mass of a few MeV and lifetime
in the range 10-50 seconds has been advocated in
ref. \cite{turn}.

An MeV tau neutrino, though cosmologically interesting,
does not obviously fit with the data on solar and
atmospheric neutrinos \cite{Smirnov}, for which neutrino
oscillations with quite small mass differences provide the most
plausible solutions \cite{valle,DARK92B}. However, so far
all attempts \cite{ma1,moh} to obtain an MeV
$\nu_{\tau}$ with the lifetime required to revive the
cold dark matter picture have ignored solar as well as
atmospheric neutrino data.
It is desirable to develop a coherent model which not
only fits COBE and IRAS observations, but also provides
solutions to the solar and atmospheric neutrino problems.

In this note we realize the Mev tau neutrino hypothesis
in a model that can naturally reconcile the cosmological
data on primordial density fluctuations with an
explanation of the solar and atmospheric neutrino
deficits through neutrino oscillations. The simplest
way to do this is to assume the few MeV $\nu_{\tau}$
to be a majorana neutrino and not a Dirac \neu as
assumed in \cite{turn}. Indeed, an Mev \nt can not
pair up with $\nu_{\mu}$ or with $\nu_{e}$ in order
to form a Dirac state, because of the laboratory bounds
on the masses of $\nu_{\mu}$ or $\nu_{e}$. As a result,
an MeV Dirac state would be obtained only by pairing off
the two-component \nt with a sterile neutrino state of
the same mass. In this case at least three other light
\neu species, one of which should also be sterile,
would be required in order to fit together the data
on solar and atmospheric \neu oscillations. This
follows from the fact that in this case the oscillations
of $\nu_{\mu}$ or $\nu_{e}$ into the MeV \nt cannot solve
the atmospheric or the solar neutrino problems and, on
the other hand, the $\nu_e-\nu_{\mu}$ oscillations could
solve either but not both, since they require quite
different values for the corresponding (mass)$^2$
differences \cite{valle}.
\ignore{
Moreover, an Mev \nt may be only marginally allowed by
energy loss considerations in supernovae ???
Indeed a supernova would rapidly loose its energy through
emission of the sterile component unless trapping ...
}

Hence, the most economical way to reconcile COBE and IRAS
observations with the cold dark matter picture
and with solar and atmospheric \neu data should
involve the presence of just one very light sterile
neutrino and just one two-component MeV state: the
majorana tau neutrino.

To construct a consistent and appealing theoretical
model is a non trivial task. Apart from naturally
relating the required smallness of the sterile \neu
mass to a suitable symmetry, one has to obey
a number of constraints:

\noindent ({\em i})
The mass and mixing of the very light sterile neutrino
\ns should not conflict with the constraints coming
from the nucleosynthesis \cite{Barbieri} and supernovae
\cite{Peltoniemi}.

\noindent ({\em ii})
The model should contain three different mass scales
namely $m_{\nu_\tau} \sim MeV$ to account for the tau neutrino mass,
$\Delta_{S}\sim 10^{-6}$ eV $^2$ or $10^{-10}$ eV $^2$ in order
to account for the solar neutrino deficit and $\Delta_{A}\sim 10^{-2}$
eV $^2$ to solve the atmospheric neutrino problem.

\noindent ({\em iii})
The $\nu_{\tau}$ should decay with lifetime in
the range $10^3-10^8$ seconds \cite{ma1} or $10-50$
seconds \cite{turn}. In the former case, it should couple
strongly to majoron, while in the latter case its decay
products should produce $\nu_{e}$.

\noindent ({\em iv})
The couplings of the majoron should be strong enough to
satisfy ({\em iii}), but this coupling should not result
in excessive energy loss through majoron emission inside
a supernova \cite{choi}.

We present below a model which successfully meets all these
requirements and discuss its most salient features.

\section{The model}

Our model is based on the triplet plus singlet majoron scheme
\cite{774} and contains three right handed neutrinos, one
of which is kept light by the imposed global symmetry.
This way of keeping the sterile neutrino light has been
already used in variety of models which tried to accommodate
the possible existence of a 17 keV neutrino state \cite{17kev}
or which try to solve the solar, atmospheric and the dark
matter problems simultaneously \cite{valle,caldwell}
\footnote{In principle, these problems can be solved
without invoking a light sterile state if neutrinos
are almost degenerate \cite{caldwell,DEG}.
However, in our present case one is obliged to introduce
a light sterile neutrino if the mass of $\nu_{\tau}$ is
to lie in the MeV range.}.

We replace the lepton number symmetry of the original
singlet majoron model by a generation dependent global
symmetry $U(1)_G$ under which the various fields transform
in the manner shown in table 1. The generation dependent
symmetry serves two purposes. It keeps the sterile $\nu_e^c$
light and it leads to the decay of the tau neutrino into
lighter neutrinos plus a majoron \cite{V}. The $SU(2)\times
U(1)\times U(1)_G$ invariant couplings of the neutrinos are given by:
\bea
\label{yuk}
{\cal L}_Y & =&  \frac{1}{2}
\frac{m}{<T^0>} \nu_{\tau L}^T C \nu_{\tau L} T^0  - \frac{\phi^0}{<\phi^0>}
     \left[m_1 \bar{\nu}_{eL} \nu_{\mu R}+
     m_2 \bar{\nu}_{\mu L} \nu_{\mu R}+ m_3 \bar{\nu}_{\tau L}
     \nu_{\tau R} \right]  \nonumber \\
& & + \frac{\mu}{<\sigma>} \nu_{eR}^T C \nu_{\tau R} \sigma
+ \frac{1}{2} M \left[ \nu_{\mu R}^T C \nu_{\tau R}
+ \nu_{\tau R}^T C \nu_{\mu R} \right]\;\;+H.c.
\label{lag}
\eea

The above Yukawa interactions lead to the following neutrino masses:
\begin{equation}
{\cal L}_m=\frac{1}{2}{\nu_L}^T \: C \: M_{\nu} \: \nu_L \: +H.C.
\end{equation}
where $M_{\nu}$ is a $6\times 6$ matrix having the following
form in the left-handed basis
$\nu \equiv (\nu_{e}^c,\nu_{e},\nu_{\mu}, \nu_{\tau},
\nu_{\mu }^c,\nu_{\tau }^c)^T$:
\begin{equation}
M_{\nu}=\left(
\begin{array}{cccccc}
0&0&0&0&0&\mu\\
0&0&0&0&m_1&0\\
0&0&0&0&m_2&0\\
0&0&0&m_{\nu_{\tau}}&0&m_3\\
0&m_1&m_2&0&0&M\\
\mu&0&0&m_3&M&0\\
\end{array}  \right)
\label{MAT}
\end{equation}

We assume the bare mass $M$ to be much greater than other
scales appearing in \eq{MAT}, as in the seesaw model. These
two heavy states can be diagonalized out leading to four light
states, one of which is massless, from the form of \eq{MAT}.
The effective 4 $\times$ 4 \neu mass matrix obtained in
the seesaw approximation takes the following form:
\begin{equation}
m_{eff}= \left(
         \begin{array}{cccc}
0&\beta \cos\theta&\beta \sin\theta&0\\
\beta \cos\theta&0&0&\alpha  \cos\theta\\
\beta \sin\theta&0&0&\alpha \sin\theta\\
0&\alpha \cos\theta&\alpha \sin\theta & m_{\nu_{\tau}}\\
\end{array}
\right)\end{equation}
where,
\begin{equation}
\beta^2\equiv(\frac{\mu}{M})^2(m_1^2+m_2^2)\;\;\;\;\;
\alpha^2\equiv(\frac{m_3}{M})^2(m_1^2+m_2^2);
\end{equation}
and
\begin{equation}
\tan\theta\equiv\frac{m_2}{m_1}  \nonumber
\end{equation}
Note that the $\nu_{e R}$ is not allowed to receive
a large mass and is kept light after the seesaw mechanism.
As already mentioned, one of the four light states described
by $m_{eff}$ is in fact massless, $\lambda_1=0$.
The other three are massive with eigenvalues
$\lambda_{2,3,4}$ approximately given by:
\begin{eqnarray}
\lambda_2 & \approx & - \beta+ \frac{\alpha^2}{2m_{\nu_\tau} } \\\nonumber
\lambda_3 & \approx &  \beta+\frac{\alpha^2}{2m_{\nu_\tau} }\\
\lambda_4 &\approx& m_{\nu_\tau} - \frac{\alpha^2}{m_{\nu_\tau} }
\end{eqnarray}
These eigenvalues nicely reproduce the hierarchical
scales required for our purposes. Because of the chosen
quantum numbers with respect to $U(1)_G$, the $\nu_{\tau}$
is the only state to receive the mass from the $SU(2)$
triplet fields and could be in the MeV range. The
parameters $\alpha$ and $\beta$ could be much smaller
than \mnt if the seesaw mass scale $M$ is chosen
appropriately large. In this case, the matrix $m_{eff}$
itself has a seesaw structure. The effective matrix describing
the mixing of $\nu_{e,\mu}$ and the sterile neutrino
($\nu_{s} \equiv \nu_{eR}$) is easily seen to posses
an approximate $L_s-L_{\mu}-L_e$ symmetry. This leads
to a pair of almost degenerate states $\lambda_{2,3}$.
Their mass provides the (mass)$^2$ difference
\begin{equation}
\Delta_A \equiv \lambda_2^2-\lambda_1^2 \approx \beta^2
\end{equation}
The approximate symmetry and hence degeneracy among two of the
neutrinos is broken by the corrections $\O(\frac{\alpha^2}{m_{\nu_\tau}})$
arising out of the vacuum expectation value (VEV) of the $\sigma$ field
which breaks the global symmetry and generates the majoron.
These corrections are naturally small because of the second
seesaw mechanism and they provide another (mass)$^2$ difference
\begin{equation}
\Delta_S \equiv \lambda_2^2-\lambda_3^2 \approx \frac{2\beta
\alpha^2}{m_{\nu_\tau} }
\end{equation}
For the values $\alpha \sim \beta\ll m_{\nu_\tau}$ one naturally obtains
the hierarchy $\Delta_S\ll \Delta_A$.

The mixing among the four light states is specified by
the matrix:
\begin{equation}
U \; m_{eff} \; U^T = diag. (0,m_{\nu_2},m_{\nu_3},m_{\nu_4})
\end{equation}
where $m_{\nu_i} \equiv \abs{\lambda_i}$
and $U$ is defined by
\begin{equation}
U^T\sim \left(
         \begin{array}{cccc}
0&\frac{1}{\sqrt{2}} & \frac{i}{\sqrt{2}}&\frac{\beta\alpha}{m_{\nu_\tau}^2}\\
-\sin\theta&\frac{1}{\sqrt{2}}\cos\theta&-\frac{i}{\sqrt{2}}\cos\theta&
-\frac{\alpha}{m_{\nu_\tau} }\cos\theta\\
\cos\theta&\frac{1}{\sqrt{2}}\sin\theta&-\frac{i}{\sqrt{2}}\sin\theta&
-\frac{\alpha}{m_{\nu_\tau} }\sin\theta\\
0&\frac{\alpha}{\sqrt{2} m_{\nu_\tau} }&-\frac{i\alpha}{\sqrt{2}
m_{\nu_\tau} }&1\\
\end{array}
\right)
\end{equation}
The weak eigenstates are related to the mass eigenstates as follows:
\begin{eqnarray}
\nu_s&  \approx
&\frac{1}{\sqrt{2}}(\nu_{2L}+i\nu_{3L})+\frac{\beta\alpha}{m_{\nu_\tau}^2}
\nu_{4L} \\\nonumber
\nu_{eL}& \approx &\cos\theta \frac{1}{\sqrt{2}}(\nu_{2L}-i\nu_{3L})
-\frac{\alpha}{m_{\nu_\tau}}\cos \theta\nu_{4L}-\sin\theta\nu_{1L} \\\nonumber
\nu_{\mu L}& \approx &\sin\theta \frac{1}{\sqrt{2}}(\nu_{2L}-i\nu_{3L})
-\frac{\alpha}{m_{\nu_\tau}}\sin \theta\nu_{4L}+\cos\theta\nu_{1L} \nonumber\\
\nu_{\tau L}& \approx &\nu_{4L}+\frac{\alpha}{\sqrt{2}m_{\nu_\tau}}
(\nu_{2L}-i\nu_{3L})
\end{eqnarray}
This gives rise to the following pattern for neutrino oscillations.
Consider first the limit in which $\Delta_{S}$ is neglected
in comparison to $\Delta_{A}$. In this limit, there are no
oscillations involving the sterile state \ns. On the other
hand the \ne and \nm oscillate among themselves with the
probability:
\begin{equation}
P_{e\mu}=\sin^2\;2\theta\;\;\sin^2\;\left(\frac{\Delta_{A}t}{4E}\right)
\end{equation}
These oscillations could give rise to an explanation of
the observed depletion in the atmospheric neutrino flux
if the parameters lie in the range \cite{Smirnov,valle}
\begin{equation}
\sin^2\;2\theta= 0.35-0.8 \:;  \: \: \: \:
\Delta_{A}=(0.3-2) \times 10^{-2}  \:
\rm{eV^2}
\end{equation}
When the smaller (mass)$^2$ difference $\Delta_{S}$ is turned on,
the \ne and the \nm start oscillating into sterile state \ns.
The probabilities for the $\nu_e$ oscillations averaged over
the shorter atmospheric \neu oscillation scale set by $\Delta_{A}$
is given by
\begin{eqnarray}
P_{ee} (t) & = & \frac{c^4}{2}\left(1+\cos(\frac{\Delta_{S}t}{2E})\right)+s^4
\nonumber\\
P_{e\mu} (t) & = & \frac{c^2s^2}{2}\left(3+\cos(\frac{\Delta_{S}t}{2E})\right)
\nonumber\\
P_{es} (t) & = & \frac{c^2}{2}\left(1-\cos(\frac{\Delta_{S}t}{2E})\right)
\end{eqnarray}
Since the mixing angle involving $\nu_{2}$ and $\nu_{3}$
is 45$^o$, the nonadiabatic MSW solution \cite{MSW} to the
solar neutrino problem is not feasible in the present case.
However, the relevant mass scale $\Delta_{S}$ could be
naturally very small so that the solar neutrino flux may
get depleted through the long wavelength vacuum oscillations
\cite{vac}. For example, $m_{\nu_{\tau}}=5$ MeV,
$\beta= 0.1$ eV and $\alpha= 0.05$ eV would give
$\Delta_{S} = 10^{-10} eV^2$ and
$\Delta_{A} = 10^{-2} \rm{eV^2}$.
One sees that similar values of $\alpha$ and $\beta$
naturally result in a very large hierarchy between the
solar and atmospheric scales. On the other hand, if one
chooses the Dirac masses $m_{1,2,3}$ in the GeV range,
then the required values of $\alpha,\beta$ can be obtained
by choosing the bare mass $M$ for the right handed neutrino
around the intermediate scale $10^{11}$ GeV. One concludes
that the required values of $\Delta_{A}$ and $\Delta_{S}$
do follow for natural choices of the parameters.

Note that, although it has been shown that \ne to \ns
oscillations where \ns is a sterile state are disfavored
by the combined Homestake and Kamiokande data \cite{petcov},
in the present case the situation is more complex, since not
only the sterile state \ns but also $\nu_{\mu}$
are involved. Since both relevant mixing angles are
large
\footnote{A possible restriction on large angle \neu
mixing has been argued to follow from the observed
energy spectra of $\bar{\nu}_e$ from supernova SN1987A
\cite{spergel}. However, ref. \cite{spergel} considered
only the simplest case of two flavour \ne to \nm mixing,
whereas in the present case one has three \neu species
one of which is sterile, so that those arguments do not
directly apply.}, a phenomenologically consistent
solution should exist.
In order to determine its parameters more sharply a more
detailed analysis of the existing solar \neu data for the
present case where both \nm and \ns take part in the solar
\neu oscillations would certainly be desirable.

\section{Cosmology}

Let us now turn to the cosmological aspects of the model.
A stringent constraint on this scenario comes from primordial
big bang nucleosynthesis, which requires that the effective
number of degrees of freedom $g_{eff}(T)$ contributing to
the energy density of the universe at the nucleosynthesis
time ($\sim$ 1 second) be less than 11.3 \cite{nucl1}.
Here $g_{eff}(T)$ has been defined in terms of the total
energy density as $\rho\equiv \frac{\pi^2}{30} g_{eff}(T)
T^4$ which includes the contribution of the relativistic
species as well as that of the nonrelativistic $\nu_{\tau}$.
The latter could violate this bound if it had only the
conventional weak interactions. However, the presence
of the majoron in our model alters the situation.
The majoron in the model is easily seen to be
\begin{equation}
\label{prof}
J \approx \sigma_I - \frac{2 \omega}{u}T_I + \frac{4 \omega^2}{u v}\phi_I
\end{equation}
where $u,v,$ and $\omega$ denote ($\sqrt{2}$ times)
the vacuum expectation values of the $\sigma,\phi^0$ and
$T^0$ scalar multiplets, respectively, while the suffix
$I$ denotes the corresponding imaginary parts. Note that
because of the hierarchy $\omega \sim MeV\ll u\sim 100 GeV$,
the invisible decay of $Z$ to the majoron is enormously
suppressed, in accordance with LEP data \cite{LEP1}, unlike
in the purely triplet majoron scheme.
The $\nu_\tau$ couples to the majoron dominantly through
its triplet admixture. Using \eq{yuk} and \eq{prof} this
coupling may be given as follows
\begin{equation}
\label{Jnn}
{\cal L}_J \approx -i\: J
\left\{ \frac{m_{\nu_\tau}}{u} \nu_{4 L}^T C \nu_{4 L} \: + \:
\frac{\alpha}{\sqrt{2} u}
\left[ \nu_{4L}^T C (\nu_{2L}-i\nu_{3L})
+ (\nu_{2L}-i\nu_{3L})^T C \nu_{4L} \right] \right\}
+ \: H.c.
\end{equation}
The $\nu_{\tau}\sim \nu_4$ has a diagonal coupling to the majoron
given by $g_J \equiv \frac{m_{\nu_\tau}}{u}$. For sufficiently
large $g_J$ the contribution of $\nu_{\tau}$ at the time of
nucleosynthesis can be suppressed through \nt annihilation.
In order for this to happen, one requires \cite{ma1}
\begin{equation}
T_{EQ1} \equiv \frac{4}{3}\;m_{\nu_\tau} Y < 0.13\;\; MeV
\end{equation}
which corresponds to $g_{eff}(T\sim 1MeV)<11.3$. Here $Y$ determines
the abundance of the nonrelativistic $\nu_{\tau}$. Using the
standard expression \cite{KT} for $Y$ we obtain
\begin{equation}
\label{TEQ1}
T_{EQ1}\approx (2.66 \times 10^{-2}
MeV)\;\left(\frac{10\;MeV}{m_{\nu}}\right)^2
         \;\left(\frac{u}{100\;GeV}\right)^4\;
\left(\frac{x_f}{4}\right)^2
\end{equation}
where
\begin{equation}
x_f \equiv \ln z - (n+1/2) \ln \ln z
\end{equation}
and
\beq
 z = 0.038 (n+1) g/g_*^{1/2} M_{Pl}m_{\nu_{\tau}} \sigma_0
\eeq
where $n=1$, $M_{Pl}=1.2 \: \times 10^{19}$ GeV is the Planck mass,
$g=2$ and, in our case \cite{ma1}, $g_*=10$. Finally,
$\sigma_0=\frac{m_{\nu_{\tau}}^2}{2 \pi u^4}$ denotes the
\nt annihilation cross section.

Physically, $T_{EQ1}$ represents the temperature  at which the
contribution of the $\nu_{\tau}$ starts dominating over that of
radiation. As would be expected, this should happen after the
nucleosynthesis epoch. After $T_{EQ1}$, the $\nu_{\tau}$ would
dominate the energy density of the universe until it decays.
This decay  would make the universe radiation dominated and
this could serve to delay the time at which matter starts
dominating again. If one fixes the overall scale of the power
spectrum of the density fluctuations in the cold dark matter
scenario from the large scale  COBE observations, then one can
also fit the small scale density fluctuations provided one
chooses $g_{eff}(T_{EQ2}) \approx \frac{25}{9}\: \times 3.36$
\cite{ma1,turn}.
Here $g_{eff}(T_{EQ2})$ determines the total contribution in
radiation at the temperature  $T_{EQ2}$ when the radiation
and matter densities again become equal. The  $\nu_e,\nu_{\mu}$,
majoron and photons contribute the amount 3.17 to $g_{eff}(T_{EQ2})$.
The remaining must come from the decay products of the $\nu_{\tau}$.
This can be translated \cite{ma1}
into a constraint on the $\nu_{\tau}$ lifetime:
\begin{equation}
\tau_{\nu_{\tau}}\approx 7.2 \times 10^{19} \:
\left(\frac{T^0}{T_{EQ1}}\right)^2 \rm{sec}
\end {equation}
where $T^0=2.73$ is the present temperature of the universe.
Since $T_{EQ1}<0.13$ MeV from nucleosynthesis,
  $\tau_{\nu_{\tau}}>2\times 10^2$ sec. On the other hand,
$T_{EQ1}$ should be higher than the temp $T_{EQ2}\sim$ few eV of
the conventional matter radiation equality, giving an upper bound
$\tau_{\nu_{\tau}}<10^{11}$ sec. In fact, requiring that the \nt
contribution to the present total energy density of the universe
is not too large gives a more stringent upper bound
on $\tau_{\mu_\tau}$ as a function of the mass \mnt \cite{KT,RHO1}.
For example, for \mnt $\sim 1$ MeV, $\tau_{\nu_{\tau}} \lsim 10^{8}$ sec.

The above considerations show that a $\tau_{\nu_{\tau}}$
in the range $\sim 10^{2}-10^{8}$ seconds would be able
to delay matter radiation equality without conflicting with
the nucleosynthesis picture.

We now show how the lifetime required by cosmology can naturally
occur in the present model. For this note that \eq{Jnn} gives rise
to the decay of $\nu_{\tau}$ to the lighter states $\nu_{2,3}$
plus a majoron, with decay width given by
\begin{equation}
\Gamma_{\nu_{\tau}}=\frac{m_{\nu_{\tau}}}{4 \pi}
\left(\frac{\alpha}{u}\right)^2
\end{equation}
Using the expressions for $\Delta_A$ and $\Delta_S$
this can be written as
\begin{equation}
\label{taunt}
\tau_{\nu_{\tau}}=(1.6 \times 10^3 sec.)\; \left
(\frac{m_{\nu_{\tau}}}{10\;MeV}\right)^{-2}\;\;
\left(\frac{u}{100 GeV}\right)^2\;\;
\left(\frac{\Delta_{A}}{10^{-2} eV^2}\right)^{1/2}\;\;
\left(\frac{10^{-10} eV^2}{\Delta_{S}}\right)\;\;
\end{equation}
It follows from the above equation that one can
simultaneously accommodate the solar and atmospheric
neutrino deficits and also have a tau neutrino decay
with the required lifetime if the singlet VEV $u$ is
chosen around $50-200$ GeV. In Fig.1 we display the
region of \nt mass and the singlet VEV $u$ allowed by
the various constraints. The solid curves a and b correspond
to $g_{eff}$ = 10.86 and 11.3, respectively. The regions to
the right of these curves would be disallowed by nucleosynthesis.
For illustration, we give curves c and d corresponding to \nt
lifetimes $10^{3}$ and $10^{4}$ seconds, respectively.

Another constraint on the model comes from the supernova. The MeV
tau neutrino with relatively large couplings to the majoron may cause
supernova to rapidly loose energy through majoron emission. The
relevant constraints have been looked at in detail in ref. \cite{choi}
in the case of the simplest singlet majoron model. In our case the
relevant coupling is the first in \eq{Jnn}. Since, this coupling
is similar (apart from a factor 2) to that of the singlet majoron
model we can adopt the analysis of ref. \cite{choi}. In our case,
the following values of $u$ and \mnt are seen to be disallowed
\begin{equation}
5.7 \times 10^{-3} < \:
\left(\frac{m_{\nu_{\tau}}}{10 \;MeV} \right) \:
\left(\frac{100 GeV}{u}\right)^2\;<\;0.82
\end{equation}
It follows from the above two equations
that it is indeed possible to obtain
the $\nu_{\tau}$ lifetime in the required range and solve the
solar and atmospheric neutrino problem for parameter choices
lying outside the range forbidden by the supernova.

The time structure of the SN87A antineutrino pulse
may also be used to constrain the lifetime of $\nu_{\tau}$.
The decay of a massive $\nu_{\tau}$ emitted in the supernova
on the way could lead to a delayed signal in the detectors.
The absence of such a signal has been used in \cite{asm}
and subsequently in \cite{pg} to put an upper bound on
the $\nu_{\tau}$ lifetime. A rough estimate of the resulting
lifetime has been given in ref. \cite{moh} to be $\sim$ 300
seconds for an Mev $\nu_{\tau}$ decaying by majoron emission.
While this is consistent with the bound we have obtained of
$2\times 10^2$ seconds, there is considerable room to relax
this supernova bound. Firstly, only a fraction $\cos^2 \theta$
of $\nu_{\tau}$ decay to $\nu_{e}$ in the present case.
Moreover, original $\nu_{\tau}$ flux may also be suppressed
by the Boltzman factor for larger masses
$m_{\nu_{\tau}} > 10 MeV$. These could result in considerable
weakening of the upper bound. Finally, in the derivation
of the supernova bound one would have to be careful in
differentiating between \nt and $\bar{\nu}_\tau$ decays to
$\bar{\nu}_e$.

The presence of a light sterile neutrino mixing with ordinary
neutrinos is in general constrained by nucleosynthesis
\cite{Barbieri}. Neutrino oscillations in the early universe
would bring the sterile \neus into equilibrium at the time of
nucleosynthesis with the active ones.  The relevant oscillation
scale in the present case is  $\Delta_{S} \sim 10^{-10} \rm{eV^2}$.
The corresponding wavelength is too large to equilibrate the sterile
neutrinos in the early universe. In fact the constraints on the
relevant mass scale is trivially satisfied in our case.

\section{Discussion}

We have attempted in this paper to provide a coherent explanation of
quite distinct phenomena in neutrino physics. The example given here is
able to resurrect the cold dark matter picture of structure formation
by making it consistent with both COBE and IRAS observations of
primordial desnsity fluctuations through a $\nu_{\tau}$
of few MeV mass and lifetime in the range $10^2-10^4$ sec. It also explains
the solar and atmospheric neutrino problem in terms of mixing among three
light neutrinos \ne, \nm and \ns.
The scheme presented here is not unique but is certainly most economical
from the point of view of explaining various phenomena mentioned above.
More importantly, it contains predictions which can be tested in the
near future in long baseline \neu oscillation experiments, as
well as some reactor \neu experiments. Moreover, due to the
presence of the weak scale majoron, the present model
allows for a distinctive signature of the invisibly decaying Higgs boson
$h \ra JJ$ \cite{JoshipuraValle92}, which could substantially affect
higgs boson search strategies at LEP \cite{alfonso},
NLC \cite{EE500}, as well as LHC \cite{granada}.

Finally, notice that, since $\nu_{\tau}$ is much heavier than other
neutrinos and has very small mixing with them
(\O($\alpha/m_{\nu_\tau} \sim 10^{-7}$))
there are no oscillations involving the \nt and experiments
such as CHORUS and NOMAD \cite{chorus} looking for the
$\nu_{\mu}-\nu_{\tau}$ oscillations should show negative
result. This feature is a generic  expectation in any model
having $\nu_{\tau}$ in the MeV range. It also implies a
strong suppression in the neutrinoless double beta decay rate.

\noi
{\bf Acknowledgements}
This work was supported by DGICYT under grant number
PB92-0084 and by the sabbatical grant SAB94-0014 (A.S.J.).
We thank R. Ghandi and P. Krastev for fruitful correspondence.

\newpage
\noi

\begin{table}
\begin{center}
\begin{math}
\begin{array}{|c|crrr|} \hline
 &  SU(2) & Y & G & \\[0.2cm]
\hline
\ell_{Le, \mu}  &  2 & -1 & -1 & \\[0.1cm]
\ell_{L\tau}  &  2 & -1 & 1 & \\[0.1cm]
e_{R}, \mu_{R}      &  1 & -2 & -1 & \\[0.1cm]
\tau_{R}   &  1 & -2 & 1 & \\[0.1cm]
\nu_{eR}   &  1 & 0 & -2 & \\[0.1cm]
\nu_{\mu R}   &  1 & 0 & -1 & \\[0.1cm]
\nu_{\tau R}   &  1 & 0 & 1 & \\[0.1cm]
\hline
\phi &  2 & -1 & 0 & \\[0.1cm]
T & 3 & 2 & -2 & \\[0.1cm]
\sigma & 0 & 0 & 1 & \\[0.1cm]
\hline
\end{array}
\end{math}
\end{center}
\caption{$SU(2) \ot U(1)_Y \ot U(1)_G$ assignments of the
leptons and Higgs scalars. Quarks are $U(1)_G$ singlets.}
\end{table}
\newpage
\noi
{\bf Figure Caption}
\noi

Fig.1 shows the region of \nt mass and lifetimes
as a function of the singlet VEV $u$ which are allowed by
the various constraints. The solid curves a and b illustrate
the nucleosynthesis contraints corresponding to $g_{eff}$ = 10.86
and 11.3, respectively. The regions to the right of these curves
would be forbidden. Curves c and d correspond to \nt lifetimes
$10^{3}$ and $10^{4}$ seconds, respectively.

\end{document}